\documentstyle[sprocl]{article}
\newcommand{\bm}[1]{\mbox{\boldmath $#1$}}     

\def\bea{\begin{eqnarray}}
\def\eea{\end{eqnarray}}

\begin{document}

\title{ON THE (NON) EXISTENCE OF SEVERAL GRAVITOMAGNETIC EFFECTS }
\author{J.-F. Pascual-S\'anchez}
\address{Dept. Matem\'atica Aplicada Fundamental, 
Secci\'on Facultad de Ciencias,
Universidad de Valladolid, 47011, Valladolid,\\ Spain\\ 
E-mail: jfpascua@maf.uva.es}

\maketitle
\abstracts{Due to the resemblance between Maxwell and the gravitomagnetic equations
obtained in the weak field and slow motion limit of General Relativity,
 one can ask if it is possible to amplify a seed intrinsic rotation or 
 spin motion by a gravitomagetic dynamo, in analogy with the well-known
 dynamo effect.
Using the Galilean limits of the gravitomagnetic equations, the answer 
to this question is negative, due to the fact that a "magnetic" Galilean
limit for the gravitomagnetic equations is physically inconsistent.
Also, we prove that, in spite of some claims, a gravitational Meissner effect does not exists}

\section{Gravitomagnetic and Maxwell equations}

Our starting point is the resemblance between Maxwell-Lorentz's electromagnetic equations and the linear and slow motion aproximation of the Einstein's equations of General Relativity.
Hence, I do not start from the full non-linear Einstein's equations, to develop, after the projection into the local rest spaces of a congruence of observers, the Maxwell analogy in General Relativity, based on the correspondence between the Faraday tensor of the electromagnetic
field and the Weyl tensor of the gravitational tidal field. This analogy has been developed in several recent papers, however 
it was put forward and clearly exposed in  ~\cite{bel1}
~\cite{pen} ~\cite{ja}.
This approach is done without any aproximation and, in this framework, the Bianchi identities are dynamical and the Einstein equations can be interpreted  as constitutive relations of a 4-dim non-linear elastic medium, (this can be seen in ~\cite{yo}).

In Newtonian theory of gravity, no fundamental gravitational force
is associated with the rotation of a mass. In this theory, if a body rotates,
the gravitational force it exerts on other masses, changes only to the 
extent that the matter distribution within the body is affected by the rotation.
The Newtonian gravitational force is only associated to the distribution of
mass at a time, but not with the state of intrinsic rotation of this mass.\\
However, Lense and Thirring (1918) and Thirring (1921) showed that, a certain
gravitomagnetic field is indeed associated with the rotation of a mass,
in the framework of the weak field aproximation to General Relativity.
%\end{equation}
From the linearized Einstein's equations, one obtains, when the first order effects of the motion of the sources are taken into account, the following Maxwell-like ({\it gravitomagnetic}) equations, which can be considered invariant under the Poincar\'e (or even Conformal) group:
\begin{eqnarray}
\nabla \bm{g}             & =& -4\pi\, \rho,   \label{13}\\
\nabla\wedge \bm{b} &=& -4\pi\,\rho\,\bm{u}+							 \displaystyle\frac{\partial {\bm{g}}}{\partial t}, \label{14} \\
\nabla\wedge \bm{g} &= &-\displaystyle\frac{\partial \bm{b}}{\partial t}, \label{15}\\
\nabla \bm{b}            &=&0. 	\label{16} 
\end{eqnarray}
Where \bm{g} is the Newtonian gravitostatic field with source the density of mass-energy, $\rho$, and \bm{b} is the gravitomagnetic field with source the density of mass current generated by the motion, in particular, an intrinsic rotation.
This deduction can be seen, in the corresponding chapters of several
%work of H. Peng G.R.G. 15 (1983) 725, with some changes in the notation
books, see for instance
~\cite{oh}, with some changes in the notation and new symbols.
Moreover, for a stationary field one obtains an
equation for $\bm{b}$, that is analogous to the electromagnetic one, changing the magnetic dipole moment by minus twice the spin angular momentum $\bm{S}$.
\section{Magnetic dynamo theory}
The term "dynamo effect" in magnetohydrodynamics (hereafter MHD), is generically used to describe the systematic and sustained generation of magnetic energy as a result of the stretching action of a velocity field
$\bm{u}$, on a magnetic field $\bm{B}$. In other words, if a conducting fluid moves in a magnetic field $\bm{B}$, the flow will be affected by the force due to the interaction between $\bm{B}$ and the currents of the fluid. Also, $\bm{B}$ will be modified (amplified) by the currents of the fluid and this is the dynamo effect.

%The full equations of MHD consist in:\\
%1.- Euler or Navier-Stokes for the fluid.\\
%2.- Conservation of mass for the fluid.\\
%3.- Conservation of momentum of the fluid and momentum of the %electromagnetic field (EM), jointly.\\
%4.- Conservation of total energy of fluid plus EM field.\\
%5.- Maxwell equations in the macroscopic magnetic-Galilean %limit.\\
%6.- Ohm's law, relating the electric current $\bm{J}$ to the %electric and magnetic fields.\\
The kinematic dynamo is the most simple case of self-excited one, due to the fact that the back reaction of the magnetic field to $\bm{u}$ is assumed negligible, and considers the evolution (amplification) of magnetic field according the induction equation:
\begin{equation}\label{23}
\frac{\partial \bm{B}}{\partial t}=\nabla\wedge(\bm{u}\wedge \bm{B})+\frac{1}{4\pi}\,\eta_e\, \Delta \bm{B},
\end{equation}
being $\eta_e$, the resistivity or difussivity (for insulators is infinite, for plasmas is zero), the reciprocal of the electric conductivity $\sigma$.
The induction equation is obtained from the "macroscopic" magnetic Galilean limit (will be discussed) of Maxwell's equations, in which case the displacement current is neglected, and Ohm's law. 
I will try to propose a similar mechanism in gravitomagnetism, to amplify $\bm{b}$ and hence the intrinsic angular momentum $\bm{S}$, due to the fact that we have Maxwell-like equations for gravity at our disposal.
However, the key equation of the kinematic magnetic dynamo, which stablish 
the loop to amplify $\bm{B}$ is the Ohm's law. Do we have a similar equation in gravity?

\section{An analog for gravitomagnetism of the Ohm's law}
Our main radical and new idea is that, in order to have a gravitomagnetic dynamo, the source fluid can not be a perfect fluid. The fluid must be "not dry", wet, and hence must have viscosity. But, as viscosity is a tensorial object and as we need a scalar, we only consider its trace, the viscous pressure, neglecting the shear viscosity. Viscosity (viscous pressure) $\eta$, will be the analog in gravitomagnetism of the resistivity $\eta_e$, for a conducting electrical medium.
Our Ohm's-like law for the moving viscous fluid, in a moving frame, will be:
\begin{equation}\label{99}
\bm{j}=\rho\, \bm{u}=\delta\left(\bm{g}+\bm{u}\wedge \bm{b}\right),
\end{equation}
where $\bm{j}$ is the mass current that appears in the first term of the r.h.s. of (\ref{14}) and being $\delta=1/\eta$, the "dryness'' of the viscous fluid.
With the Ohm's like law (\ref{99}) and following the same procedure as in electromagnetism, one obtains an induction equation for $\bm{b}$.
The difference with (\ref{23}) is a change of sign in the second term of the r.h.s., i.e., the presence of a "concentration" term instead of a diffusion one. This gives rise to some problems concerning the existence of a gravitomagnetic dynamo, however, in the next section I present a stronger reason against it.

\section{Galilean limits of the gravitomagnetic equations}
It is well-known that Maxwell's equations and the Lorentz force law have two different kinds of Galilean limits: electric and magnetic.
This is due, from the mathematical point of view, to the existence of two different kinds of Galilean four-vectors.
Starting from a Lorentz four-vector, for instance $(\bm{E},\bm{B})$, this can be more timelike, i.e.,  $| \bm{E} |  >\!\!>| \bm{B} |$, and in this electric Galilean limit, its transformation under the Galilean inertial one is:
\begin{equation}\label{101}
\bm{E}^\prime=\bm{E},\qquad \bm{B}^\prime=\bm{B}-\bm{v}\wedge \bm{E}.
\end{equation}
Physically, in the electric limit one describes situations where isolated electrical charges move at low velocities. On the other hand, the magnetic Galilean limit (in which the space-like parts are dominant) is the usual situation at the macroscopic level where magnetic effects are dominant, due to the balance between negative and positive electric charges.
This magnetic Galilean limit is the proper one that is used in magnetic dynamo theory but it is not possible in gravitomagnetism, where we do not have negative masses at our disposal. Thus, in gravitomagnetism, if we take a Galilean limit, this must necessarily be of the electric (almost Newtonian) kind and describe situations where isolated masses move at low velocities.
In this electric (almost Newtonian) limit, the gravitomagnetic equations (\ref{13},\ref{14} and \ref{16}) have the same expressions, but there is an important difference, in this limit the Faraday-like equation (\ref{15}) has not induction term, this equation now read
\begin{equation} \label{150}
\nabla\wedge \bm{g} =  0,
\end{equation}
Moreover, {\it in this limit, the proper one for gravity, it is impossible to build a gravitomagnetic dynamo even if we use an Ohm's like law for gravity as (\ref{99})}, because we do not have, at our disposal, an induction term in the Faraday's equation. 

The only possibility that remains, in my opinion, 
to construct a gravitomagnetic dynamo, would be to consider 
a non-relativistic generalized Newtonian theory of gravity  
of the kind introduced 
by Bel in ~\cite{bel3}. This possibility will be explored in a future work.

\section{The gravitomagnetic Meissner effect does not exists}
Working with the gravitomagnetic equations, several works have appeared (see for instance ~\cite{la}), in which a gravitational analog of the electromagnetic Meissner-Ochsenfeld effect is presented. 
As a result, Lano, by using the classical London equations, suggests an expulsion of the gravitomagnetic field $\bm{b}$ from the core of the neutron stars to the exterior, i.e., a transport outwards of the spin angular momentum, due to the diamagnetic nature of the Meissner effect. 
I will show that this gravitomagnetic Meissner effect does not exists in gravitomagnetism and instead of a diamagnetic nature, at the classical level in spite of the fact that is a truly quantum effect, one finds a paramagnetic character. Surprisingly, this key difference comes from a trivial, but fundamental, error in the calculation of ~\cite{la}. 
Begin with the Lorentz force law in the electric (quasi-Newtonian) limit, 
$\partial \bm{j}/\partial t=\rho\, \bm{g}$,
that is the first London type equation. By substituing it into the Faraday's law, (\ref{15}), one obtains
\begin{equation} \label{151}
\frac{\partial }{\partial t}\left(\frac{1}{\rho}\nabla\wedge \bm{j}+\bm{b}\right)=0.
\end{equation}
One solution is the second London equation
%\begin{equation}
$\nabla\wedge\bm{j}= -\rho\,\bm{b}$.
%\end{equation}
From the Ampere equation in the magnetic limit,  
\begin{equation} \label{152}
\nabla\wedge\bm{b}=-4\pi\bm{j},
\end{equation}
taking the curl, substituing the second London equation
%\[\nabla\wedge(\nabla\wedge\bm{b})= 4\pi\bm{\rho}\,\bm{b},\]
and finally appling (\ref{16}), one obtains:
\begin{equation} \label{153}
	\Delta \bm{b} = -4\pi\bm{\rho}\,\bm{b}.	
\end{equation}
So, we have found a "paramagnetic" character instead of the Meissner-Ochsen\-
feld effect. Our final criticism to our previous deduction is similar to the proposed for the dynamo effect. In this case one uses a mixture of Galilean limits (electric, for the Lorentz force and magnetic, for the field equations), thus the paramagnetic character of gravitomagnetism must be also put into question.

\section{Conclusion}
Our final remark is the following. The use  of the linear slow motion aproximation of Einstein's equations can lead to the appearance of spurious effects. The truly gravitomagnetic effects must be found using the Maxwell analogy of General Relativity, first exposed by Ll. Bel, i.e., working with tidal curvature fields instead of with kinematic connection
fields.

\section*{Acknowledgments}
I am grateful to A. San Miguel and F. Vicente for discussions and TeX help, to A. Ferriz-Mas and M. N\'u\~nez for introducing me to the subject of magnetic dynamo theory
and, after the completion of this paper, to Ll. Bel for drawing my attention to one of his publications. This work has been partially supported by the spanish research projects VA61/98 of Junta de Castilla y Le\'on and C.I.C.Y.T. PB97-0487.

%}
\end{document}